# Dose and energy dependence of mechanical properties of focused electron beam induced pillar deposits from $Cu(C_5HF_6O_2)_2$

V Friedli [1], I Utke [1], K Mølhave [2] and J Michler [1]

[1] Empa, Swiss Federal Laboratories for Materials Testing and Research, Laboratory for Mechanics of Materials and Nanostructures, Feuerwerkerstr. 39, 3602 Thun, Switzerland
[2] Technical University of Denmark, Department of Micro- and Nanotechnology, Bldg. 345e, 2800 Kgs. Lyngby, Denmark

E-mail: ivo.utke@empa.ch

**Abstract**
Bending and vibration tests performed inside the scanning electron microscope were used to mechanically characterize high-aspect pillars grown by focused electron-beam (FEB) induced deposition from the precursor $Cu(C_5HF_6O_2)_2$. Supported by finite element (FE) analysis the Young's modulus was determined from load-deflection measurements using cantilever-based force sensing and the material density from additional resonance vibration analysis. The pillar material consisted of a carbonaceous (C, O, F, H containing) matrix which embeds 5...10 at. % Cu deposited at 5 keV and 20 keV primary electron energy and 100 pA beam current, depending on primary electron energy. Young's moduli of the FEB deposits increased from 17±6 GPa to 25±8 GPa with increasing electron dose. The density of the carbonaceous matrix shows a dependence on the primary electron energy: 1.2±0.3 g cm$^{-3}$ (5 keV) and 2.2±0.5 g cm$^{-3}$ (20 keV). At a given primary energy a correlation with the irradiation dose is found. Quality factors determined from the phase relation at resonance of the fundamental pillar vibration mode were in the range of 150 to 600 and correlated to the deposited irradiation energy.

## 1. Introduction
The increasing importance of gas-assisted focused electron-beam (FEB) induced deposition as an extremely flexible, maskless direct-write nanofabrication technique for three-dimensional prototyping at sub-100-nm scale demands in-depth studies of the mechanical properties of such obtained deposits. Such studies will be useful to progress in rapid prototyping of nanowire or nanotube hybrid structures which requires in situ bonding after correct placement in an electron microscope [1-5]. In addition FEB fabricated scanning probe microscopy sensors, like magnetic [6], thermal [7], optical [8, 9], and atomic force sensors [10-12], as well as grippers [13] and nanoimprint tools [14] require analysis of their mechanical properties and their dependence on deposition parameters in order for further mechanical optimization. Mechanical studies are also important for small deposits obtained from gas assisted focused ion beam (FIB) induced deposition such as grippers [15] and biocell surgery [16]. In contrast to gas assisted FIB deposition, FEB induced deposition proceeds without contamination (no ion implantation into the growing deposit) and with better resolution. The smallest deposit dimensions so far obtained with FEB were 1 nm dot deposits [17] and 5 nm for freestanding pillars [18].

FEB (and FIB) deposits obtained from carbon containing organometallic molecules feature a composite structure, *i.e.* nanocrystalline metal grains embedded in a carbonaceous matrix [19]. Such a composite structure is often advantageous for sensors in terms of magnetic, optic, or electric functionality (which is supplied by the dispersed metallic nanocrystals) and simultaneous mechanical and chemical stability supplied by the carbonaceous matrix. Evidently, the mechanical properties, such as stiffness, density, strength, toughness, and adhesion must be determined experimentally since no data is available from template bulk materials. Both, static and dynamic mechanical test methods provide an effective way to characterize materials at the micro- and nanomechanical level [20]. For one-dimensional pillar-like nanostructures, including nanowires and nanotubes, measurement techniques based on bending and vibration tests are well suited. Of note, in mechanics the term pillar often relates to a structure under compressive load. In this paper the structures are subjected to bending and vibration, however, we adhere to the term pillar as the FEB and FIB literatures refers to this term when a high-aspect ratio structure is deposited coaxially within a finely focused electron beam. The integration of such mechanical test setups in a scanning electron microscope (SEM) has proven to be a successful approach to allow for visual observation of tests on individual nanostructures [21]. Furthermore, the SEM is the ideal instrument to determine the three-dimensional geometry of the tested structures which is a crucial task to obtain accurate mechanical properties.

Density and Young's modulus measurements of FEB or FIB deposited pillars are still very rare (FIB: [22-27], FEB: [28-30]) and the precision of the methodology not yet quantitatively determined. In the following we report on bending and vibration tests inside a SEM to measure the force constant, resonance frequency, and, for the first time, the quality factor of pillars deposited by FEB induced deposition from the organometallic precursor $(Cu(C_5HF_6O_2)_2 \cdot xH_2O$, Copper[II] hexafluoroacetylacetonate (hfa) hydrate, CAS: 155640-85-0). Furthermore, the values of Young's modulus and the density of the deposited materials were extracted by finite element (FE) analysis and we discuss the influence of non-cylindrical pillar geometries. We compare our results to available literature data and discuss the dependence of irradiation dose, deposited energy, and primary electron energy on the mechanical properties of the pillar deposits.

## 2. Theory

For single clamped cylindrical pillars with elastically isotropic material properties, uniform diameter, $d$, and length, $l$, mechanical theory (Euler-Bernoulli beam model for bending deflection, which neglects shearing and rotary inertia) provides a formulation of the Young's modulus in terms of the pillar's force constant, $k_d$, as [31]

$$E = \frac{64}{3\pi} \frac{l^3}{d^4} k_d. \quad (1)$$

In addition the velocity of sound in the material relates to the resonance frequency of the fundamental flexural mode, $f_d$, as [32]

$$\sqrt{\frac{E}{\rho}} = 7.148 \frac{l^2}{d} f_d. \quad (2)$$

Deviations from the ideal cylindrical shape will considerably alter the pre-factor. For instance, for conical pillars (2) changes to $\sqrt{E/\rho} = 2.883 \cdot l^2/d_{\text{base}} \cdot f_d$ [33], corresponding to a 5x reduction of $\sqrt{E/\rho}$ for a conical pillar with diameter $d_{\text{base}} = 2d$ compared to a uniform and otherwise equivalent pillar. This highlights that the shape variation of the width must be correctly input into the model and the dimensions accurately measured. Nonaka et al. used an analytical model based on a polynomial fit of the diameter versus length variation of non-uniform pillars and solved the free vibration equation by the variation principle [34]. We used static and dynamic FE analysis to model the load-deflection and vibration behaviour of cylindrical pillars with non-uniform diameter.

## 3. Experiment
### 3.1. Sample preparation

For mechanical characterization of FEB deposited materials, pillars with aspect ratios >30 were deposited from the precursor $Cu(C_5HF_6O_2)_2$ (see figure 1) in a SEM equipped with a thermal tungsten emitter and a homebuilt gas injection system (GIS). The FEB irradiated the substrate at normal incidence and the pillars grew vertically into the stationary beam. All depositions were performed on a Si substrate with a native oxide at a probe current of 100 pA measured in a Faraday cup and at an acceleration voltage of 5 kV and 20 kV. The 5 kV series in figure 1 shows that best focus conditions gives the highest growth rate. The local precursor flux to the deposition site was determined as $1.3 \times 10^{17}$ cm$^{-2}$ s$^{-1}$ from the measured precursor mass loss rate and Monte Carlo simulations [35] of the gas flow through the tube nozzle of the GIS in the molecular flow regime. The backpressure in the vacuum chamber during the deposition experiments was $3 \times 10^{-5}$ mbar.

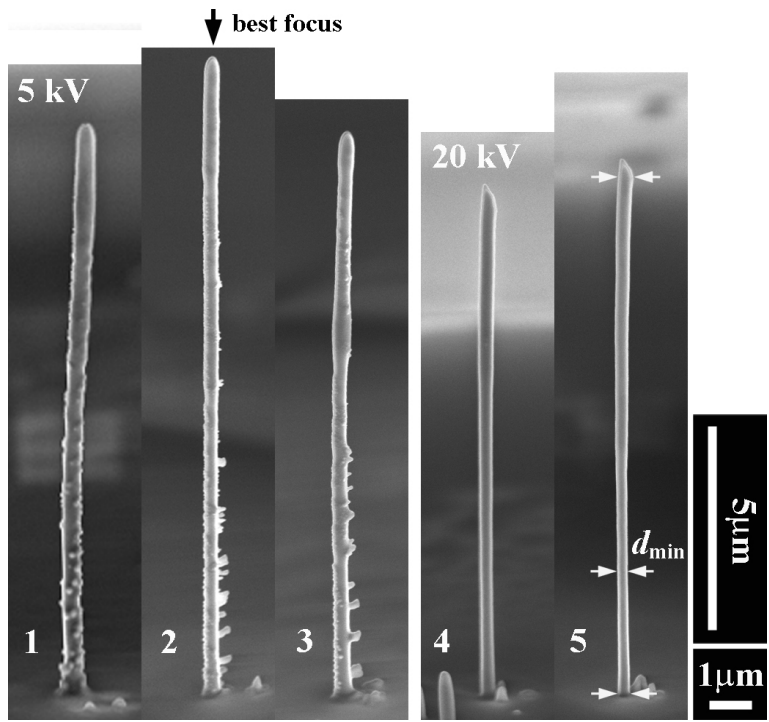

**Figure 1:** FEB deposits grown normal to the substrate in continuous spot mode for 30 minutes. SEM observation with 68° substrate tilt. Deposition conditions: 100 pA probe current, acceleration voltage is indicated. Precursor: $Cu(C_5HF_6O_2)_2$. Note the non-uniform diameter of the pillars 3, 4 and 5. For pillar 5 the variation of the base and the top diameter relative to the indicated minimal diameter, $d_{min}$ = 238 nm, is +15% and +36%, respectively.

The pillars were grown at >10 µm lateral pitch distance on the substrate which avoided proximity effects [36] and additional deposition [37]. It has been experimentally observed that slight bending is induced during observation of the pillars in a field-emission gun SEM probably due to surface stress by a deposited contamination layer [38]. The vertical deposition rate, averaged over the entire deposition time, was in the order of 7-9 nm s$^{-1}$ at both 5 and 20 kV. Although the deposition time was 30 min for all pillars, a difference in deposit height is observed which is attributed to focus adjustments between subsequent depositions. The 5 kV-pillars show side roughness whereas the pillars grown at 20 kV have smooth surfaces. This phenomenon has been reported in the literature several times but is currently not fully understood [39]. The influence of this roughness on the mechanical measurements was considered to be negligible.

### 3.2. SEM integrated bending tests

SEM integrated bending experiments for the determination of stiffness relies on physical interaction between the pillar and a micromachined reference cantilever used as the force sensor. We used a SEM integrated nanomanipulation setup composed of two vacuum compatible positioning units. For the approach towards the pillar sample, the cantilever was mounted on a three-axes nanomanipulator with two rotational and one linear degree of freedom (MM3A, Kleindiek Nanotechnik, Reutlingen, Germany). The sample was mounted on a three-axes Cartesian nanopositioning stage with integrated capacitive position sensors (P-620 series, Physik Instrumente (PI), Karlsruhe, Germany) which provide close-loop positioning with sub-nanometre resolution. The entire setup was mounted on the SEM stage such that the pillar axis was at an angle of 85° with the FEB.

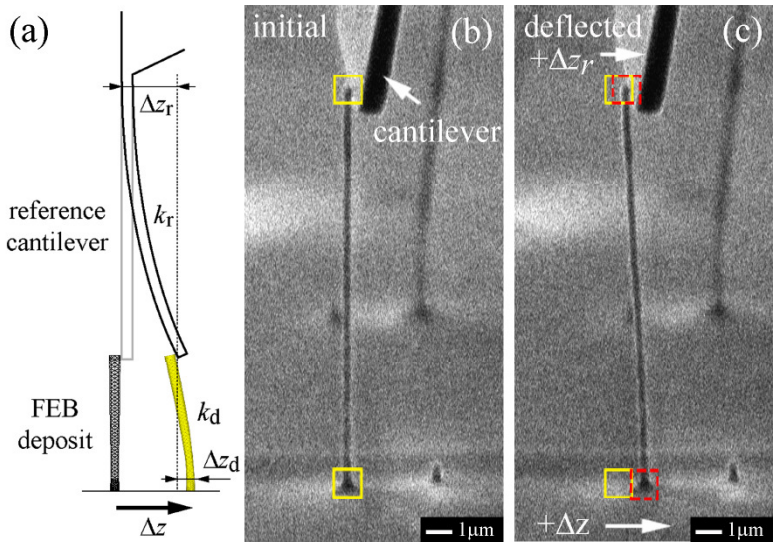

**Figure 2:** (a) SEM integrated bending test on a FEB pillar-like deposit using a reference cantilever as the force sensor (not to scale). The applied force at the contact point of cantilever (index r) and deposit (index d) reads $F = k_r \Delta z_r = k_d \Delta z_d$, with force constant $k$ and deflection $z$. $k_d$ is determined by measuring the deposit and cantilever deflection upon a stage displacement of $\Delta z$. Note the non-uniform pillar shape (exaggerated for better visibility) used for FEM analysis. (b) SEM image of a bending test on pillar 2 (see figure 1) in initial unstrained contact position marked by solid boxes. (c) Pillar deflection upon a horizontal shift of the substrate by $\Delta z = 1\mu m$. The pillar top and root are marked by dashed boxes as tracked by an auto-correlation image processing software.

In a bending test as depicted schematically in figure 2 (a) the load-deflection relation at the contact position is given by

$$F = k_r \Delta z_r = k_d \Delta z_d, \qquad (3)$$

where $F$ is the force component acting perpendicular to the sample axis. $k_r$ and $k_d$ are the force constants and $\Delta z_r$, $\Delta z_d$ the deflections of the reference cantilever and the sample, respectively. For the experiments a reference cantilever was chosen which had a force constant close to the force constant of the pillar samples. This $k$-matching is required since the measurement accuracy decreases if either the sample or the reference cantilever deflects only marginally upon loading [40]. The reference was a microfabricated cantilever from $SiO_2$ with length $l_r = 200$ μm and width $w_r = 10$ μm. The fundamental resonance frequency of the cantilever was determined by SEM imaging of the maximum vibration amplitude at 21.88 ± 0.01 kHz. The reference force constant $k_r = 59.311 w_r (l_r f_r)^3 \sqrt{\rho_{SiO_2}^3 / E_{SiO_2}}$ evaluated to 0.019 N m$^{-1}$ suffering from a maximum systematic error of 0.004 N m$^{-1}$ (22%), assuming a SiO$_2$ density of 2.2 g cm$^{-3}$ and Young's modulus of 70 GPa [41] and variations in thermally grown thin film SiO$_2$ properties of $\Delta E_{SiO2} = 0.1 E_{SiO2}$, and $\Delta \rho_{SiO2} = 0.03 \rho_{SiO2}$. The corresponding cantilever thickness is found as $h_r = 1.787 l_r^2 \sqrt{12 \rho_{SiO_2} / E_{SiO_2}} f_r = 0.96\,\mu m$, which is close to the nominal value of 1 µm.

A bending test on one of the pillars is shown in figure 2 (b) and (c). Generally, pushing the sample against the force sensor (+Δz direction) deflected both the pillar and the cantilever. In all experiments the deflection Δz$_d$ was kept <10% of the pillar length. Additional experiments confirmed that the bent pillars flexed back to their initial unstrained position upon release of the bending load by retracting the reference cantilever and thus no plastic deformation occurred. The quantification of bending tests was performed based on image post-processing of video data captured by SEM imaging during the tests. A cross-correlation algorithm detected the location of a previously defined image detail and the corresponding coordinates were saved in the video time line sequence. Both, the root and the pillar location were traced using this technique and revealed i) the relative stage displacement Δz and ii) the pillar deflection Δz$_d$ = Δz - Δz$_r$ during the bending test experiment. The data of one experiment presented in figure 3 shows the measured pillar deflection in terms of the applied load F. The negative load corresponds to a retraction of the pillar from the cantilever beyond their unstrained position. In this retraction regime the pillar-cantilever contact is maintained by attractive electrostatic forces induced during the observation with the electron beam. The reproducibility of iterative bending measurements on the same pillar was very high which confirms that deflections $< 0.1 \Delta z_d / l$ did not influence the stiffness of the material and supports that the experiments are performed within the elastic deformation limit. The displacement noise observed in the post-processed deflection data originating from noisy SEM images was maximally one pixel which corresponded to a position accuracy of 20 nm. With the reference cantilever used a force resolution of 0.4 nN was obtained.

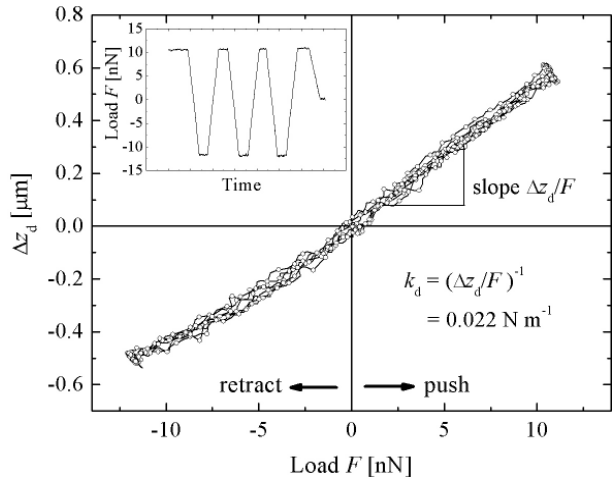

**Figure 3:** Pillar deflection Δz$_d$ in terms of the applied load F from SEM images of a bending test on pillar 1 (see figure 1). The data was extracted by a cross-correlation image processing software. The inset shows the corresponding sequence of sample deflection cycles. The deflection response at positive loads (cantilever is pushed against the pillar) is linearly fitted and reveals the force constant of the pillar sample. At negative loads cantilever and pillar remain in contact by electrostatic attraction.

The pillar force constant, $k_d$, is determined using (3). The tilt angle between the cantilever and the pillar sample was kept minimal to minimize errors from this source [2]. In some cases we observed non-linear deflection behaviour in the pillar-cantilever retraction regime. We speculate that such a behaviour results from a slightly bi-stable behaviour of the cantilever due to its particular very slightly curved shape which results in a non-linear force constant at retraction loads larger than about -2 nN from its unstrained position at $F$ = 0 nN (see figure 3). Of note is that the results of the deflection test are dominantly influenced by the properties of the electron-impact dissociated base material where the highest strain occurs [42].

*3.3. SEM integrated vibration tests*

In a vibration test the pillar samples were mechanically excited at their fundamental resonance mode as shown in figure 4. The vibration amplitude maximum which determines the resonance frequency, $f_d$, was detected in two ways inside the SEM: the overall modal shape and the absolute maximum deflection amplitude for a given excitation amplitude was visualized by SEM imaging and the frequency spectrum was taken by the secondary electron (SE) signal through the interaction of a stationary beam with the vibrating structure [23]. If the stationary electron beam irradiates the maximum amplitude position, a peak in the integrated SE signal is detected at resonance while sweeping the excitation frequency. This is due to the increasing dwell time of the vibrating sample inside the beam. Best results were achieved by slightly defocusing the stationary electron beam which increased the dynamic range of the technique due to a spatially increased interaction between the beam and the vibrating pillar. Phase locking of the time-resolved SE and excitation signal allows extracting the response of amplitude *and* phase at resonance from the noisy SE signal. The position of the peak maximum corresponds to the resonance frequency $f_0$. It should be noted that the acquired SE signal due to cantilever deflection is in general not linear with the cantilever's vibration amplitude. Hence, errors are introduced when measuring the frequency peak width and determining the quality factor with the formula $Q_A = \sqrt{2} f_0 / FWHM$. Calculating the quality factor of the pillar vibrations using the phase slope at resonance $Q_P = f_d / 2 |d\phi/df|_{f_d}$ avoids such errors. For example, in figure 4 (b) the error is $Q_A / Q_P = 1.3$.

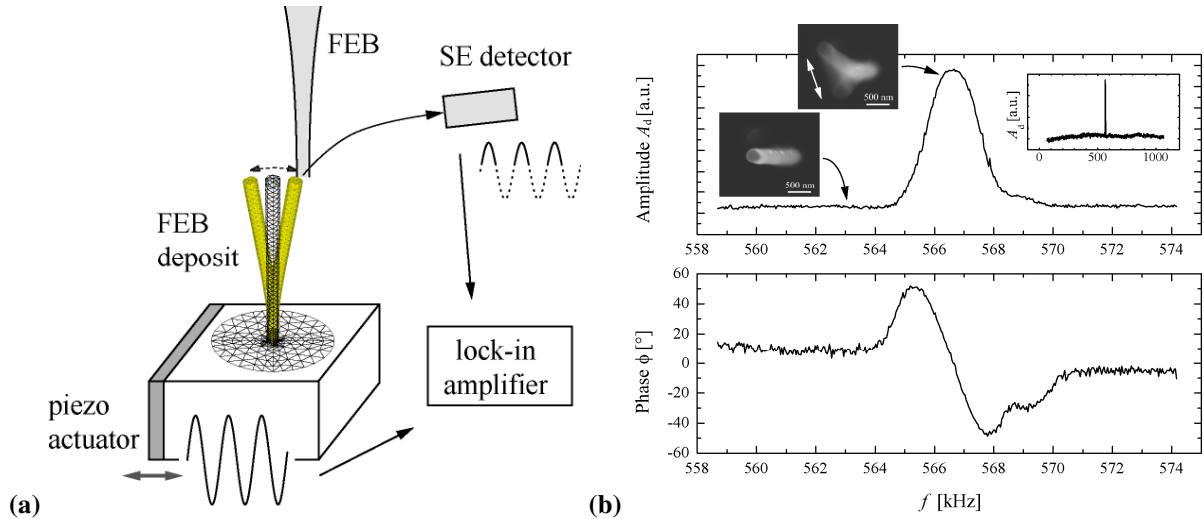

**Figure 4:** (a) Measurement principle in a piezo-driven resonance experiment based on the stationary beam technique. Locking the sample deflection signal from the secondary electron (SE) detector to the excitation signal reveals the deflection amplitude $A_d(f)$ and the phase characteristics $\phi(f)$ at resonance. (b) Spectrum of the fundamental resonance of pillar 5. Insets: Top-view SEM images show the slightly tilted pillar excited at off-resonance and at resonance. The SE deflection signal is locked to the excitation in the range between 565..568 kHz. The inset on the right shows a full range spectrum up to 1 MHz of the amplitude response which locates the peak at $f_0$ = 566.6 kHz. The phase slope at resonance corresponds to a quality factor of 274.

A typical resonance test is shown in figure 4 (b). In a first step an overview spectrum was acquired with the stationary beam technique to roughly locate resonance peaks by sweeping the excitation frequency through the full available detection bandwidth of up to 1.2 MHz. Secondly, close-up spectra were acquired at the frequencies of interest. In all the resonance test experiments the excitation amplitude was adjusted to limit the deflection amplitude peak to < 10 % of the pillar length to avoid plastic deformation.

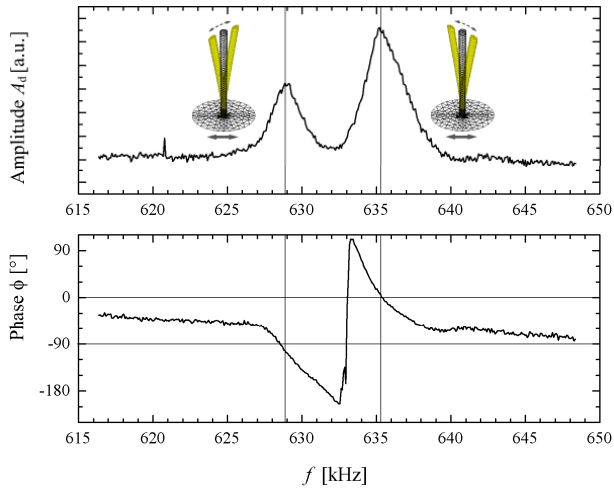

**Figure 5:** The stationary beam technique detects polarized vibrations as a double amplitude peak and a 90° phase-shift between the two resonances in the secondary electron spectrum. The insets illustrate the corresponding orthogonal vibration directions.

In some of the experiments we observed two resonance modes vibrating along the orthogonal principal axes. To detect all resonance modes, spectra were routinely acquired along two perpendicular directions with the stationary beam technique at top-view incidence. Once the orthogonal directions of the polarized resonance modes were identified the stationary electron beam was positioned at 45° relative to these directions. This allowed detecting all peaks in a single frequency sweep as shown in figure 5. If two peaks were observed we used the average frequency to evaluate (2).

Orthogonal resonance modes probe the geometrical deviations from the ideal straight cylinder shape. They can be attributed to both a non-circular pillar cross-section and a curved pillar shape [43]. Theory predicts two orthogonal resonance modes for a pillar with elliptical cross section; the frequency ratio being proportional to the ratio of the two principle diameters. From our FE simulations we found that the intrinsic curvature of a pillar, modeled as a torus segment, splits up the resonance into an in-curvature-plane mode and a mode normal-to-curvature-plane. The FE results proved an approximation formula $f_{\text{deflected}}/f_{\text{straight}} \approx \sqrt{1 + 0.04 l^2 / R^2}$ [44], for a pillar with length $l$ and radius of curvature $R$, to be consistent to within a few %. Accordingly, a pillar which is curved by 20 % from straight at its top ($R/l = 2.5$) has an in-curvature-plane resonance frequency which is shifted by +0.32 % with respect to the straight pillar. Our FE analysis further predicts in this case a splitting of the two orthogonal frequencies by about 0.23 %.

*3.4. Finite element analysis*

The experimentally determined values of the pillar force constant, $k_d$, and resonance frequency, $f_d$, as well as the experimentally determined pillar shape were the input for FE analysis executed using the software ANSYS Multiphysics. The Young's modulus and the density were then determined such that the model calculations were adjusted to match the experimental results. The optimum mesh grid was determined by adjusting the grid density until the FE results varied less than 0.01 % when compared to a further simulation run using a finer mesh grid.

The shape of the pillars was determined from SEM images and modelled by a spline function connecting measurements of the pillar diameter at typically five positions distributed along the pillar axis such that minimum and maximum diameter values were included. Such rotational-symmetry pillar models represent a smoothed shape of the pillar, they deviated less than ±5% from the pillar shapes observed in the SEM images. Pillars were imaged from two perpendicular directions to exclude projection errors. We found that the pillars had an ellipticity of 1-2 %. Furthermore, the silicon substrate was taken into account in the FE model to such an extent that the side and bottom surfaces did not influence the results. The model assumed a homogenous density and Young's modulus along the pillar which is reasonable with respect to our EDX measurements showing no difference in pillar composition between the base and apex part of the pillar.

## 4. Results and discussion

The details of the FEB deposited pillars and the results of the determined Young's modulus and density are compiled in table 1.

**Table 1:** Summary of measurements of FEB deposited pillars with the precursor $Cu(C_5HF_6O_2)_2$ at a probe current of 100 pA (see figure 1). The indicated precision indicates the statistical error which is estimated as described in the text. $E_0$ is the primary electron energy, $l$ and $d$ are the pillar length and diameter, $k_d$ is the force constant of the pillar deposit, $E$ its Young's modulus, $\rho_d$ and $\rho_m$ its deposit and matrix density, $f_d$ its resonance frequency, and $Q$ its quality factor. $E$ and $\rho_d$ are determined from FE analysis and the accurate pillar shapes from SEM images. $D$ is the irradiation dose and $W$ the deposited energy.

| Pillar | 1 | 2 | 3 | 4 | 5 |
|---|---|---|---|---|---|
| $E_0$ [keV] | 5 | | | 20 | |
| Composition | $Cu_{0.1}M_{0.9}$ | | | $Cu_{0.06}M_{0.94}$ | |
| | $M_{0.9}=C_{0.6}O_{0.22}F_{0.08}$ | | | $M_{0.94}=C_{0.7}O_{0.2}F_{0.04}$ | |
| $l$ [μm] | 14.56 | 15.42 | 13.70 | 12.10 | 12.73 |
| $d$ [nm] | 413 | 340 | 360 | 281 | 256 |
| $k_d$ [Nm$^{-1}$] | 0.022 | 0.011 | 0.017 | 0.013 | 0.009 |
| $E$ [GPa] | 16±2 | 20±3 | 17±2 | 23±3 | 27±4 |
| $f_d$ [kHz] | 766 | 600 | 729 | 632 | 566 |
| $\rho_d$ [g cm$^{-3}$] | 2.0±0.2 | 2.2±0.2 | 2.3±0.2 | 2.8±0.2 | 2.7±0.2 |
| $\rho_m$ [g cm$^{-3}$] | 1.0±0.1 | 1.2±0.1 | 1.3±0.1 | 2.2±0.2 | 2.1±0.2 |
| $Q$ [-] | 145 | 345 | 550 | 194 | 316 |
| $D$ [e$^-$/atom] | 9.0±0.7 | 11.4±0.9 | 10.7±0.8 | 11.3±0.8 | 13.8±1.0 |
| $W$ [keV/e$^-$] | 3.2 | 3.2 | 3.4 | 2.2 | 2.0 |
| $W$ [keV/atom] | 28.9±2.2 | 36.3±2.7 | 36.4±2.7 | 24.8±1.9 | 27.5±2.1 |

*4.1. Irradiation dose and deposited energy*

In the following we would like to discuss the mechanical properties listed in table 1 as a function of irradiation dose and deposited energy. Since we measured the composition and density of the deposits we can relate these parameters directly to the number of deposited atoms. This gives a much better idea at the atomistic level, i.e. how many electrons or amount of energy a deposited atom has received during FEB deposition; and, furthermore, it is a dimensionless representation which allows for easy comparison of other materials deposited by FEB.

The dose in units of impinging electrons per deposited atom is [45]

$$D = \frac{I_p t / e_0}{V_d \rho_d N_A / M_d}, \qquad (4)$$

where $V_d$ is the deposit volume, $I_p$ the probe current, $t$ the deposition time, $e_0$ the electron charge, and $N_A$ Avogadro's constant. The molar mass of the deposit, $M_d$, is determined from EDX composition measurements (see section 4.2.). All other parameters were also measured and the standard deviation in this method is $\sigma_D = 0.07 D$. Of note, the inverse of (4) gives the deposition yield $Y_D = 1/D$ (which is proportional to the adsorbed density of non-dissociated molecules $n$ in the irradiated region and the electron impact dissociation cross section $\sigma$, $Y_D = n\sigma$.) From classical resist based e-beam lithography it is known that in a given dose range, the irradiation dose is proportional to the number of possible chemical reactions excited by electrons *inside* the deposit (or resist), such as molecule dissociation, bond formation, or electron stimulated desorption.

The energy $W$ deposited from the primary electrons inside the pillar apex was MC simulated with MOCASIM [47], which uses the Bethe stopping power approximation [19] along the electron trajectories (see figure 6). On average, 3.3 keV and 2.1 keV per incident electron were deposited inside the pillar apexes by 5 keV- and 20 keV-electrons, respectively. The 5-keV-electrons loose more energy with respect to the 20-keV-electrons due to the known $\rho Z/M \ln(E_0)/E_0$ dependence of the Bethe stopping power and the fact that 20 keV electrons are all scattered out of the pillar volume. The deposited energy in units of energy per deposited atom is obtained by multiplying (4) with $W$ and the standard deviation becomes $\sigma_W = 0.075 W$.

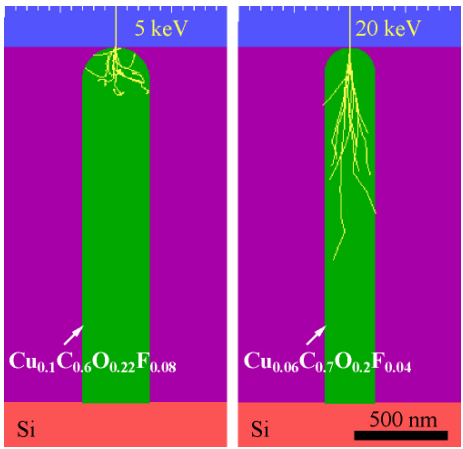

**Figure 6:** Three-dimensional Monte Carlo simulation of 10 normally impinging electrons scattered in the FEB deposited pillar. The primary electron energy is indicated. Electron trajectories outside the pillar are not drawn. The pillar material deposited at 5 keV and 20 keV is defined by its density (see table 1) and chemical composition.

With the knowledge of the deposited energy we can estimate the beam induced heating at the pillar apex. Using the formalism outlined in [19] for pillars, a maximum temperature <200°C for a thermal heat conductivity of the deposited material of 0.3 W K$^{-1}$ m$^{-1}$ is predicted. The heat conductivity value was taken for plasma-assisted CVD of low density hydrogenated amorphous carbon [48].

*4.2. Chemical composition*

The composition of the pillars measured by energy dispersive X-ray spectroscopy (EDX) is given in terms of a pseudo-binary compound $Cu_x M_{1-x}$, where M denotes the carbonaceous matrix. For example, the precursor molecule in this study $Cu(C_5HF_6O_2)_2$ is $Cu_{0.03}M_{0.97}$, where M = $(C_5HF_6O_2)_2$ =

($C_{0.36}O_{0.14}F_{0.43}H_{0.07}$). Under electron irradiation the adsorbed molecule is fragmented, volatile desorbed fragments are pumped away, and the deposit composition becomes $Cu_{0.1}M_{0.9}$ ($M_{0.9} = C_{0.6}O_{0.22}F_{0.08}$) for 5 keV-electrons and $Cu_{0.06}M_{0.94}$ ($M_{0.94} = C_{0.7}O_{0.2}F_{0.04}$) for 20 kV-electrons. The hydrogen content (which is not detected by EDX) can generally be neglected due to $x_H M_H / M_d < 1\%$, for typical hydrogen contents $x_H < 17\%$ in FEB deposits [46]. The accuracy of the measured elemental composition, used for subsequent error analysis, is within 5%. EDX spectra taken at the pillar's base and apex indicate a *homogenous* composition over the entire pillar. We attribute this to the relatively high thermal decomposition temperature of 250°C [49] for the molecule $Cu(C_5HF_6O_2)_2$ which prevents thermally activated molecule dissociation by FEB heating [19, 50, 51] during pillar growth. Of note, for molecules having thermal decomposition temperatures below 100°C compositional changes along the pillar were frequently observed: $Co_2(CO)_8$ [37, 52], $Fe(CO)_5$ [53], and (hfa)CuVTMS [54].

From table 1 it is seen that the Cu content was approximately doubled for 20 keV and tripled for 5 keV grown pillars with respect to the stoichiometry of the initially injected molecule. Interestingly, the oxygen content in the deposit is slightly larger than in the original molecule for both energies. This points to the fact that the residual gas, mainly water, inside the microscope chamber contributes to the final composition of the deposit as it was also observed for FEB induced deposition with nickel containing molecules [55]. The carbon content in the matrix is 70 at.% for the 20 keV and 60 at.% for the 5 keV pillars. Remarkably, the content of the volatile element fluorine is about a factor 2 smaller for the 20 keV compared to the 5 keV. This is consistent with our matrix density measurements in section 4.4 which show a factor 2 denser matrix for the 20 keV deposits. Evidently, there is a pronounced dependence of composition on the primary electron energy in the range of 5-20 keV. Within this range of primary electrons, normally all the electronic transitions leading to molecule dissociation (dissociative electron attachment, dissociation into neutrals, dissociative ionisation) and the related electron stimulated desorption are in constant operation. Also, the relatively small dose range of 9-14 electrons/atom does not point to a change in the deposition regime, for instance from electron-limited to precursor-limited. This leads us to propose a sputtering mechanism, which relies on direct momentum transfer, as being responsible for the fluorine reduction. Such a mechanism depends on the primary electron energy: the maximum kinetic energy $E_{max}$, which can be transferred from an electron with mass $m_e$ and energy $E_0$ in a collision with an atom inside the deposit with mass $m_a$ is [19]

$$E_{\max}(E_0) \cong 4E_0 \cdot m_e / m_a . \qquad (7)$$

Maximum transferred energies are summarized for all volatile atoms of $Cu(C_5HF_6O_2)_2$ in table 2. An atom can be sputtered from the surface when the transferred energy is larger than its surface binding energy. The corresponding sputter rate can be found in [56]. Typical displacement energies for polymers were categorized to be within 2-5 eV depending on bond strength, atomic network, and atomic weight of constituent atoms. Due to the reduced atomic network, surface binding energies are somewhat smaller than displacement energies.

**Table 2:** Maximum transferred energies from electrons with initial energy $E_0$ to the volatile atoms contained in the precursor molecule $Cu(C_5HF_6O_2)_2$.

| $E_0$ [keV] | H | O | F |
|---|---|---|---|
| 5 | 10.9 | 0.67 | 0.57 |
| 20 | 44.1 | 2.75 | 2.3 |

From table 2 it is evident that a 5 keV electron can transfer only to hydrogen energy of > 2 eV necessary for sputtering. At 20 keV however, fluorine can in principle be displaced or sputtered when accepting 2 eV as a lower limit for the threshold (displacement or sputter) energy.

*4.3. Young's modulus*
The error in the determination of Young's modulus using (1) and (3) is estimated by taking into account the measurement accuracy and precision of the deposit geometry using the SEM and the calibration of the SiO$_2$ reference cantilever used as the force sensor. Since the pillar dimensions as well as the SiO$_2$ cantilever dimensions are read-out from the same SEM, the systematic error of 3% originating from its calibration [57] partly cancels out for the displacement measurements in the deflection test. The maximum systematic error in the determination of Young's moduli becomes $\Delta E/E = \pm 19\%$. Furthermore, the statistical precision of the pillar length and diameter measured from a SEM image is estimated to 2% due to edge effects. This source leads to a standard deviation $\sigma_E = 0.14 E$ in this method.

In figure 7 (a) Young's moduli of deposits from Cu(C$_5$HF$_6$O$_2$)$_2$ are plotted versus the dose. For comparison, Young's modulus of pure copper is in the order of 130 GPa, while for hydrogenated amorphous carbon (a-C:H) values between 10-140 GPa can be found [58]. For our pillar deposits a smaller range of 15 to 30 GPa is found. Within the statistical measurement error no conclusion can be drawn whether for a given constant primary energy deposit stiffness is driven by deposited energy or dose. However, a more pronounced dependence of stiffness on dose for all deposits can be observed; the energy dependence shows a large scatter in data (see table 1). It is clear that deposits obtained from 20 keV show about 30% higher stiffness than deposits from 5 keV impinging electrons. This is in contrast to measurements on FEB pillar deposits from the precursor phenanthrene (C$_{14}$H$_{10}$); Young's modulus of these deposits decreased by roughly a factor two by increasing the acceleration voltage from 5 kV to 20 kV [29]. In a nanoindentation study on FEB thin film deposits from the precursor Paraffin (C$_{22}$H$_{46}$, C$_{24}$H$_{50}$), the Young's modulus increased by a factor two with increasing acceleration voltage from 3 kV to 20 kV [2].

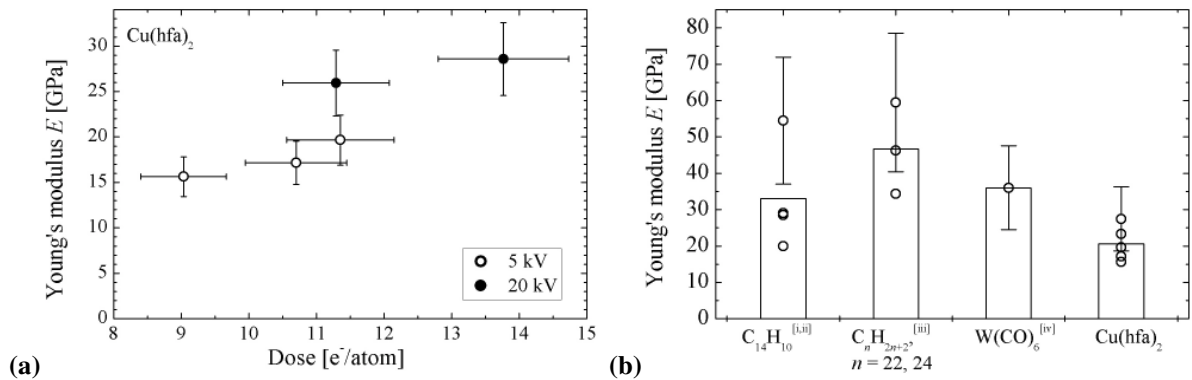

**Figure 7:** **(a)** Young's modulus versus the irradiation dose per deposited atom. The FEB acceleration voltage for deposition is indicated. **(b)** Young's modulus of FEB deposits from different precursor molecules (i: [28], ii: [29], iii: [2], iv: [59]) in comparison to our Cu(C$_5$HF$_6$O$_2$)$_2$ deposits. Bars represent the average of data points for a given molecule. The error bars in all three graphs show the standard deviation of 14% (see section 4.3). Note that in (b) the deposition parameters also change between the different molecules.

In figure 7(b) our results are compared to Young's moduli of FEB deposited materials found in the literature. If the results of differing molecules are to be compared with each other, the total error of 33% = 19% + 14% (systematic error + standard deviation) must be considered, since experiments were performed in different microscopes. The systematic error is not important when data points were collected in the same microscope and compared relative to each other as in figure 7 (a). Note that in figure 7 (b) several data points for the same precursor molecule represent measurements on materials deposited with varying conditions, e.g. acceleration voltage, beam current, and molecule supply rate. However, due to missing reported data we cannot compare this data in terms of deposited dose or energy.

*4.4. Density*

The analytical solution for the density of a cylindrical pillar shape with perfectly uniform diameter is found by combining (1) and (2):

$$\rho_d = 0.133 \frac{k_d}{l d^2 f_d^2}, \tag{6}$$

where the pillar's force constant $k_d$ is known from the bending tests (see section 3.2.) and the pillar's resonance frequency $f_d$ from vibration tests (see section 3.3.). The analysis of (6) results in a maximum systematic error of the method on the order of $\rho_d$ [g cm$^{-3}$] ± 13 [%] while the standard deviation is found to be $\sigma_{\rho_d} = 0.08 \rho_d$ assuming the same SEM measurement accuracies as for the error analysis of the Young's modulus. The determination of the resonance frequency leads to a negligible error. A rough indication of the high precision in the frequency measurement is given by the low uncertainty to determine the resonance peak maximum, given that the measured quality factors correspond to peaks with typical *FWHM* on the order of $10^{-3} f_d$. For comparison, from mass sensing experiments using a piezoresistive cantilever mass sensor integrated into the SEM, the density of FEB deposits could be determined with a maximum error on the order of only 10 % which is due to a mass sensor calibration procedure which does not rely on the material properties of the cantilever [60].

Density determination of the non-uniform diameter pillars of figure 1 was performed using FE simulations as described in section 3.4. Table 1 shows that pillars grown at 5 kV have an average density of 2.2 g cm$^{-3}$ while the pillars grown at 20 kV have a density of 2.8 g cm$^{-3}$. Interestingly, this increase in the 20-kV-deposit density is due to a considerable increase in *matrix* density and not at all due to the content of dense Cu ($\rho$ = 8.96 gcm$^{-3}$) which is in fact less than in the 5-kV-deposit. The matrix density is calculated by $\rho_m = (\rho_d - x\rho_{Cu})/(1-x)$, where $x$ is the molar copper content in the pseudobinary compound $Cu_xM_{1-x}$. We found an average matrix density $\rho_m$ = 2.2 g cm$^{-3}$ for the 20 kV pillars which is almost twice as high as the matrix density $\rho_m$ = 1.2 g cm$^{-3}$ for the 5 kV pillars.

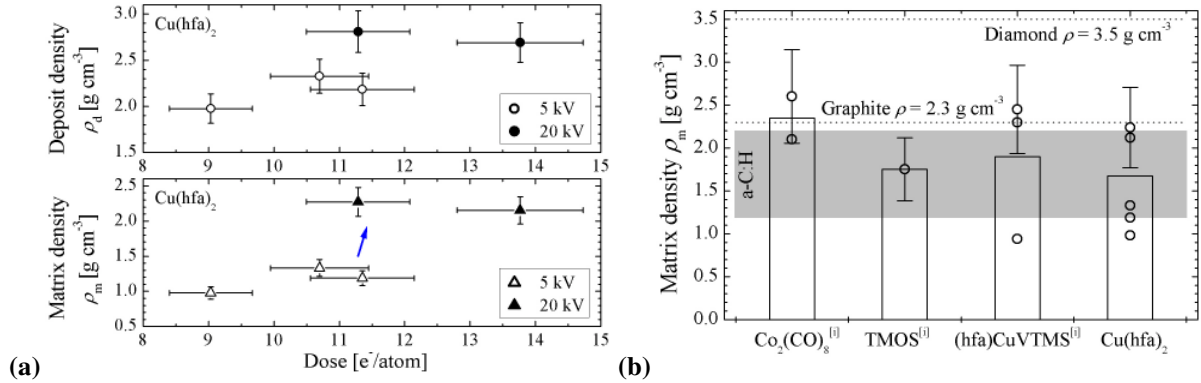

**Figure 8:** (a) Deposit density and matrix density versus the irradiation dose per deposited atom. The FEB acceleration voltage for deposition is indicated. (b) Matrix densities of FEB deposits from different precursor molecules (i: [45]) in comparison to our $Cu(C_5HF_6O_2)_2$ deposits. The error bars show the standard deviation of 8% for the deposit density and of 10% for the matrix density. Note that in (c) the deposition parameters also change between the different molecules. As a reference the a-C:H density in the range of $\rho$ = 1.2-2.2 g cm$^{-3}$ [58] deposited by plasma-enhanced CVD in different conditions and using different source gases is indicated.

As for the Young's moduli, the deposit densities show a more pronounced dependence on electron dose than on deposited energy as seen in figure 8(a). Additionally, for the matrix density a dependence on the energy of the primary electrons is seen. This can be related to the sputter mechanism discussed in section 4.2 which needs a high primary electron energy to transfer an energy to the volatile atoms higher than the displacement/surface binding energy. A matrix with less volatile atoms will then better reticulate due to the remaining non-volatile, highly reactive radicals and give a higher density. Evidently, reticulation also has a dose dependence for a given energy of primary electrons as was shown in our previous studies on materials grown with similar conditions and doses from the precursor (hfa)CuVTMS where the deposit density was measured using a cantilever-based mass sensor [45]. The matrix density of 25-kV-pillar deposits was found to be on the order of 1 g cm$^{-3}$ increasing to 2.3-2.45 g cm$^{-3}$ for an irradiation dose of 9-11 and 15-22 electrons per deposited atom, respectively.

The (matrix) density values in figure 8(a) show the same increasing trend towards higher doses and similar absolute values.

Figure 8(b) compares matrix densities of differing molecules. Matrix densities are typically lower than diamond and graphite since the deposited amorphous carbonaceous matrix contains oxygen, hydrogen, and fluorine according to the elemental composition of molecules used. The matrix densities better compare to strongly hydrogenated amorphous carbon materials.

*4.5. Dissipation*

From the phase relation at resonance the quality factors $Q$ of the pillars were measured to be on the order of 150 to 600 at room temperature in vacuum (see figure 9). To our knowledge, these are the first investigations concerning dissipation in FEB deposited pillars.

The quality factor quantifies the intrinsic energy dissipation in the vibrating structure and the dissipation to the environment [61]. The latter can be neglected in SEM vacuum conditions (background pressure $< 10^{-4}$ mbar) and thus, $Q$ describes intrinsic losses, acoustic losses in the clamping interface [62-64], and surface losses [65-68]. The phase variation at resonance was used to determine the quality factor of the $Cu(C_5HF_6O_2)_2$ pillars. Successive measurements from several spectra resulted in values that varied up to 30% for the same pillar which we suspect to be within the statistical scatter in the measurement method.

It is known that quality factors of single crystal nanostructures strongly depend on the sample volume-to-surface ratio which suggests that surface losses dominate dissipation with shrinking resonator dimensions [69, 61, 70]. Due to the low variation of the surface-to-volume ratio of the investigated pillars this relation could not be confirmed from our data. Figure 9 suggests, however, that the $Q$ of the pillars is related to the deposited energy per deposited atom. The higher energy dose means basically, that the temperature at the growing tip apex was hotter which might accelerate reticulation reactions and polymerisation leading in turn to a better homogenized material which dissipates less energy by the above mentioned mechanisms. Of course, such a "homogenisation" can be accomplished also by increasing the electron dose, in the sense of completing such reactions by more electrons.

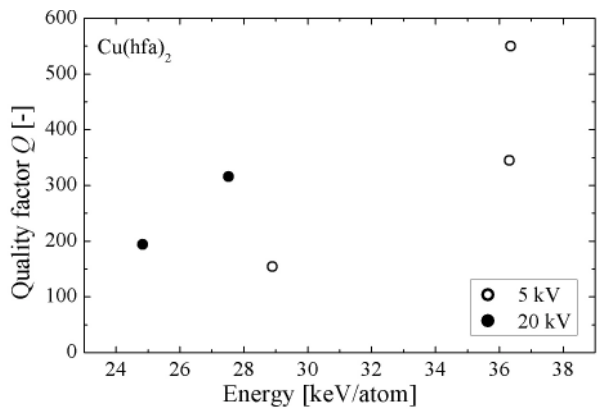

**Figure 9:** Quality factors versus the deposited energy per atom. The FEB acceleration voltage used for pillar deposition is indicated.

**5. Conclusions**

We determined the Young's modulus, density, and quality factors of individual high aspect ratio FEB deposited pillars using SEM integrated force-deflection and vibration measurements in addition to finite element analysis in order to account for the diameter variations of the pillars. We proposed a sputter mechanism to explain the lower fluorine content and the higher matrix density of the pillars deposited at 20 keV with respect to pillars deposited at 5 keV. At 20 keV all volatile atoms can be sputtered and displaced inside the carbon matrix which leads to a better reticulation and higher density. Besides the dependence on primary electron energy, an additional dependence on electron dose was found for the elastic modulus and the density for a given energy. The first quality factor measurements on such pillars favour a dependence on deposited energy. This energy was calculated via Monte Carlo simulations of electron trajectories and is a measure for the temperature in the pillar apex. Higher temperatures can increase reticulation reactions and homogenize the deposit.

**Acknowledgements**
We thank Dr. Samuel Hoffmann, EMPA, and Fredrik Östlund, EMPA, for the programming of the image processing software. Financial support is acknowledged from the EU project NanoHand.